\documentclass[a4paper,11pt]{article}
\usepackage[a4paper]{geometry}
\usepackage{times,cite}
\usepackage[scriptsize]{caption}
\usepackage{xcolor}
\usepackage{algorithm,algorithmic}
\usepackage{tikz}
\usepackage{amssymb}
\usepackage{amsthm}

\title{A simple yet efficient algorithm to turn one oriented triangular mesh connectivity into another}
      
\author{J\'er\'emy Espinas and Rapha\"elle Chaine and Pierre-Marie Gandoin}
 
 \usetikzlibrary{intersections,positioning,calc,through,decorations.pathreplacing,backgrounds}
\tikzset{
vertex/.style={circle, draw=blue!50, fill=blue!20, thick, inner sep=0pt, minimum size=1mm},
font=\tiny
}

\title{Practical Reduction of Edge Flip Sequences in Two-Dimensional Triangulations}
      
\author{J\'er\'emy Espinas and Rapha\"elle Chaine and Pierre-Marie Gandoin}

\begin{document}

%\maketitle
$  $
\vspace{1cm}

\begin{center}
\LARGE{Practical Reduction of Edge Flip Sequences in\\Two-Dimensional Triangulations}
\end{center}

\vspace{1cm}

\hspace*{0,5cm}\begin{tabular}{ccc}
\textbf{J\'er\'emy Espinas}			&\textbf{Rapha\"elle Chaine}			&\textbf{Pierre-Marie Gandoin}\\
Universit\'e de Lyon, CNRS		&Universit\'e de Lyon, CNRS			&Universit\'e de Lyon, CNRS\\
Universit\'e Lyon 1, LIRIS			&Universit\'e Lyon 1, LIRIS			&Universit\'e Lyon 2, LIRIS\\
France						&France							&France\\
\footnotesize{jeremy.espinas@liris.cnrs.fr}		&\footnotesize{raphaelle.chaine@liris.cnrs.fr}		&\scriptsize{pierre-marie.gandoin@liris.cnrs.fr}\\
\end{tabular}

\vspace{1cm}

%\maketitle

%------------------------------------------------------------------------------

\begin{abstract}
The development of laser scanning techniques has popularized the representation of 3D shapes by triangular meshes with a large number of vertices. Compression techniques dedicated to such meshes have emerged, which exploit the idea that the connectivity of a dense mesh does not deviate much from the connectivity that can be constructed automatically from the vertex positions (while  possibly being guided by additional codes). The edge flip is one of the tools that can encode the differences between two meshes, and it is important to control the length of a sequence of flips that transform one triangulation into another.
This paper provides a practical solution to this problem. Indeed, the problem of determining a minimal sequence of edge flips between two triangulations is NP-complete for some types of triangulations including manifold triangulations of surfaces, so that it is necessary to develop heuristics. Moreover, it is sometimes difficult to establish a first sequence of flips between two meshes, and we advocate a solution based on the reduction of an existing sequence.
The new approach we propose is founded on the assignment of labels to identify the edges, with a property of label transfer during a flip. This gives a meaning to the tracking of an edge in a sequence of flips and offers the exploitation of very simple combinatorial properties. All the operations are performed directly on the sequence of labels denoting the edges to be flipped, almost regardless of the underlying surface, since only local temporary connectivity is involved:
\begin{itemize}
\renewcommand{\labelitemi}{$\bullet$}
\item Given an initial triangulation $T$, and a sequence of edge flips $\phi$, we introduce three simple moves to generate new sequences $\gamma$ such that $\phi$ and $\gamma$ transform $T$ into a similar triangulation, up to a permutation of its labels. For the case where $T$ is a triangulation of a convex $n$-gon, we prove that these three moves could be used to turn $\phi$ into any other sequence $\gamma$, such that $\phi$ and $\gamma$ transform $T$ into a similar triangulation, up to a permutation of its labels.
\item On the basis of these three moves, we propose an efficient algorithm to reduce the input sequence of flips $\phi$, without ensuring the minimization of the sequence length. This algorithm is polynomial and can be used in a purely combinatorial setting as well as in a geometric setting. The effectiveness of our approach is illustrated by practical benchmarks and comments.
\end{itemize}
\end{abstract}

%------------------------------------------------------------------------------

\newpage

%\linenumbers

\section{Introduction}

In this paper, we consider triangulations of dimension $2$, possibly with boundaries, in which each face is bounded by a cycle of $3$ edges and each edge is incident to one or two faces. Besides, the vertices are assumed to be labelled. If an edge $(bc)$ is incident to faces $(abc)$ and $(dcb)$, the flip of edge $(bc)$ removes $(abc)$ and $(dcb)$ and constructs $(abd)$ and $(adc)$ which are incident to the created edge $(ad)$. There is no flip of an edge belonging to the boundary of a triangulation. The edge flipping is an operation widely used in computational geometry, especially to enumerate triangulations or to transform a mesh so that it satisfies certain properties.

The edge flip operation cannot be applied to any edge and some flips will be prohibited to maintain the structural properties of the triangulation. These structural properties may be purely combinatorial or they can also include geometric fold over free predicates determined by a particular embedding \cite{Bose09}. For example, this is the case if we restrict ourselves to planar straight-edge triangulations where each vertex has a geometric location. There is a specific case where all the edges can be flipped even if one considers geometric intersection-free properties: the triangulations of a convex $n$-gon.

Given two triangulations of the same domain with a common set of vertices, algorithms have been derived to generate a sequence of flips that transforms the first triangulation into the second. There are many ways to construct this sequence when dealing with a convex $n$-gon \cite{livre}. As far as planar straight-edge triangulations are concerned, since a sequence is reversible, it is possible to find a solution using Delaunay triangulation as a pivot between two triangulations, using the Lawson's algorithm \cite{Lawson77}. In the combinatorial case, if we restrict ourselves to triangulations of genus $0$ surfaces and impose that any two faces share at most one edge, the canonical form proposed by Wagner \cite{Wagner36} can be used in turn. Espinas's \textit{et al.} \cite{Espinas12}  work within a different class of triangulations, where any two faces can share the same triplet of vertices, and their algorithm produces a sequence of flips between two triangulations sharing the same topological genus without using any pivots.

Finding a minimal sequence of flips is more complicated. The only known method consists in building the {\it flip graph} of a given set of vertices $S$, i.e. the graph whose vertices are the possible triangulations of $S$ and where two vertices are connected if a single flip exists to transform the first triangulation into the second. Minimal flip sequences between two triangulations are the shortest paths that connect the two triangulations in the flip graph. However, it is difficult to compute these sequences since the number of triangulations grows exponentially with the number of vertices, and the flip graph quickly becomes impossible to build. However, it is known that in the case of a convex $n$-gon, the flip graph is Hamiltonian \cite{Lucas87,Hurtado99}. There are also a number of studies giving approximations of the flip distance, i.e. the number of flips in a minimal sequence. In the triangulations of a convex $n$-gon, the question of whether the determination of the flip distance in polynomial-time is still an open problem \cite{Sleator87}. However, Pournin \cite{Pournin12} showed that for a convex $n$-gon with $n \ge 13$, $2n - 10$ flips are sufficient and sometimes necessary to transform one triangulation into another. Regarding more general planar straight-edge triangulations, computing the flip distance is an NP-complete problem \cite{Lubiw12,Pilz12}, as is the case for simple polygons \cite{Aichholzer12}. However, Lawson \cite{Lawson77} showed that for any two planar triangulations consisting of $n$ vertices, $O(n^2)$ flips are sufficient, and Hurtado \textit{et al.} \cite{Hurtado96} proved that this is a lower bound for the number of flips. Hanke \textit{et al.} \cite{Hanke96} showed that the flip distance between two planar triangulations is at most the number of edge intersections obtained by superimposing the two triangulations. If the point set does not have five vertices forming an empty pentagon, Eppstein \cite{Eppstein10} showed that the flip distance can be computed in time $O(n^2)$. For the general case, he gave a polynomial-time algorithm for computing a lower bound. It can be observed that the more configurations are authorized for the flips, the better bounds are. In the combinatorial case, determining the flip distance between two triangulations of a genus $0$ surface whose flips are conditioned by the criteria proposed by Wagner \cite{Wagner36}, the best upper bound for now is $5.3n - 24.4$ edge flips \cite{Bose11}. Regarding the lower limit, pairs of triangulations exist that require $2n - 15$ edge flips \cite{Komuro97}. In this paper, we work on triangulations whose edges are \textit{labelled}. This idea has already been used for non-crossing spanning trees to which an \textit{edge move} operation is applied. This is a generalization to trees of the edge flip and it is used to modify a tree or to swap its edge labels \cite{Holzmann72,Liu88,Hernando03}. 

In this paper, we will illustrate that the labelling of edges is also advantageous when dealing with triangulations. After introducing our framework with a number of definitions and properties, we underline that some basic flip sequences correspond to transpositions of labels. We also introduce the notions of weak and strong equivalence between the flip sequences that turn a given triangulation into a target one, depending on whether or not the resulting labels are the same (section \ref{sec_Triangulation}). We then introduce three moves to generate flip sequences that are weakly equivalent to a given sequence. In the case of triangulations of a convex $n$-gon, these three moves correspond to a generating set of the equivalence classes (section \ref{sec_Generation}). These results are then used to develop an algorithm for the reduction of flip sequences which imposes no edge labelling on the initial or resulting triangulations, is adaptable to all types of triangulations and works in polynomial time (section \ref{sec_reduction}). The behaviour of this algorithm is illustrated in a practical situation.

%---------------------

\section{Sequences of edge flips in triangulations with labelled edges}
\label{sec_Triangulation}

%-------------------

\subsection{Triangulations with labelled edges}

In this work, we exploit the advantages of identifying each edge of a triangulation with a label, rather than using its endpoints, as generally done. Each edge is given a different label, which will persist throughout the edge flips, except for boundary edges that cannot be flipped. Of course, there are other cases where a labelled edge cannot be flipped, because the very definition of the flip operation is structurally conditioned by a combinatorial or a geometric intersection-free criterion. Actually, we construct the set of triangulations with labelled edges ${\mathcal T}_{lab}$ on top of a class of triangulations ${\mathcal T}$ with an edge flip operation defined on it, and the accompanying conditions required for flipping an edge. In the following, we will only impose this class of triangulations to guarantee that every face is incident to three different edges and that each edge is incident to different vertices. Then we can choose to use purely combinatorial triangulations or remain within planar triangulations, manifold triangulations of sampled surfaces or even triangulations of the convex $n$-gon.

\subsection{Flip of a labelled edge: definitions and properties}

%\subsubsection{Definitions}

Let $T \in {\mathcal T}_{lab}$ be a triangulation with a flippable edge $i$ between two vertices $a$ and $b$. $i$ is incident to the faces $(abc)$ and $(adb)$. The edge flip of $i$, denoted by ${\mathcal F}_{i}$, suppresses the edge ($ab$) and its label $i$ is transferred onto the new created edge $(dc)$. The \textit{support} of the flip ${\mathcal F}_i$ on $T$, denoted as $supp(i, T)$, consists of the two faces $(abc)$ and $(adb)$ incident to $i$ in $T$. ${\mathcal F}_{i}(T)$ denotes the triangulation that would result from the edge flip $i$ on $T$.\\

%\subsubsection{Involution:}
\hspace*{-5.4mm}\textbf{Involution:}
 If $i$ is a flippable edge in $T$, then $i$ is also flippable in ${\mathcal F}_{i} (T)$ and ${\mathcal F}_{i} ({\mathcal F}_{i} (T)) = T$, which means that ${\mathcal F}_{i}$ is an involution. \\

%\subsubsection{Sequence of labelled edge flips}

Let $T \in {\mathcal T}_{lab}$ be a triangulation with labelled edges $1, \ldots n$. We define the composition of two edge flips as $({\mathcal F}_{i_2} \circ {\mathcal F}_{i_1})(T) = {\mathcal F}_{i_2}({\mathcal F}_{i_1}(T))$, which means that it can be performed on $T$ if $i_1$ is flippable in $T$ and $i_2$ is flippable in ${\mathcal F}_{i_1}(T)$. The \textit{sequence of flips} of the edges $i_1, \ldots, i_f$ of $T$ corresponds to $\phi = ({\mathcal F}_{i_n} \circ {\mathcal F}_{i_{n-1}} \circ \ldots \circ {\mathcal F}_{i_1})$.\\

%\subsubsection{Commutativity in a sequence of edge flips}

%\subsubsection{Restricted commutativity:}
\hspace*{-5.4mm}\textbf{Restricted commutativity:}
Let $i$ and $j$ be two flippable edges of $T$, such that no face of $T$ is incident to both $i$ and $j$ (we say that the cardinality of $supp(i, T) \bigcap supp(j, T)$ is null in terms of the number of common faces and note $|supp(i, T) \bigcap supp(j, T)|=0$). We can observe that ${\mathcal F}_{i} \circ {\mathcal F}_{j}$ can be performed on $T$ and that $({\mathcal F}_{i} \circ {\mathcal F}_{j})(T)=({\mathcal F}_{j} \circ {\mathcal F}_{i})(T)$ and we say that ${\mathcal F}_{i}$ and ${\mathcal F}_{j}$ commute on $T$. 

%% TODO : REGARDER SI ON MAINTIENT LA PHRASE SUIVANTE ICI
From this, we can deduce rules of commutativity between two consecutive edge flips within a sequence: let $\phi$, $\mu$ and $\gamma$ be three sequences of flips such that $\phi (T) = (\mu \circ {\mathcal F}_{i} \circ {\mathcal F}_{j} \circ \gamma) (T)$. We have $\phi (T) = (\mu \circ {\mathcal F}_{j} \circ {\mathcal F}_{i} \circ \gamma) (T)$ if $|supp(i, \gamma(T)) \bigcap supp(j, \gamma(T))| = 0$.\\

%\subsection{Transposition of labels as a sequence of labelled edge flips}

We now introduce ${\mathcal P}_{(i,j)}$ as an operation of \textit{label transposition} between two edges $i$ and $j$ of a triangulation $T$. When being composed within a sequence, edge flips and transpositions can commute in the following way:

\[({\mathcal P}_{(i,j)} \circ F_k)(T) = (F_{{\theta}_{(i,j)}(k)} \circ {\mathcal P}_{(i,j)})(T)\] where ${\theta}_{(i,j)}(k)$ corresponds to $j$ if $k=i$, $i$ if $k=j$ and $k$ otherwise.

In fact, the main reason for introducing the transposition of labels is the relationship between this operation and specific sequences of edge flips.\\

\hspace*{-5.4mm}\textbf{Proposition 1.}
\label{prop:1}
Let $T_5$ be a triangulation of a $5$-gon in which any possible configuration of an inner edge is flippable, and let  $i$ and $j$ be the two inner edges of $T_5$. We have: $${\mathcal P}_{(i,j)}(T_5) = ({\mathcal F}_i \circ {\mathcal F}_j \circ {\mathcal F}_i \circ {\mathcal F}_j \circ {\mathcal F}_i)(T_5) = ({\mathcal F}_j \circ {\mathcal F}_i \circ {\mathcal F}_j \circ {\mathcal F}_i \circ {\mathcal F}_j)(T_5)$$

\begin{figure}[h]

\centering
\includegraphics[scale=0.5]{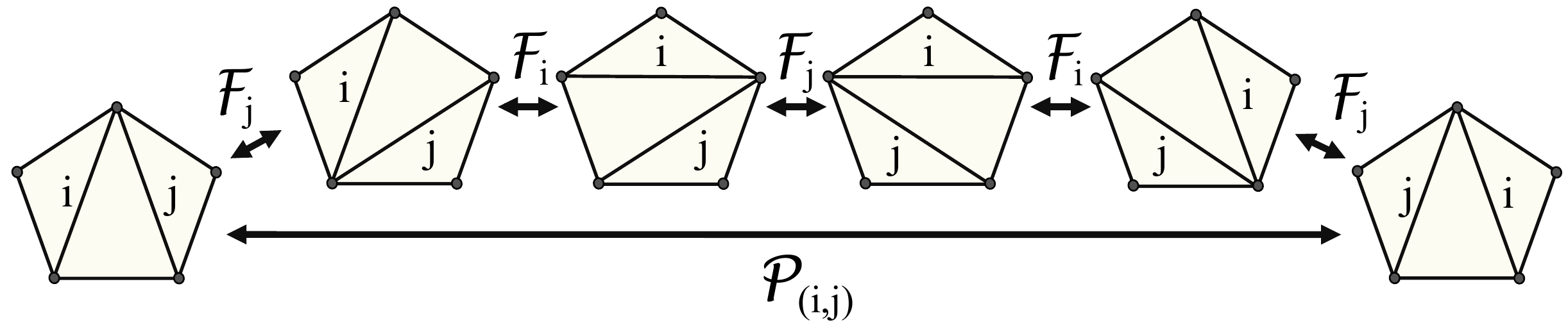}
\caption{\label{fig_Transposition}
Illustration that ${\mathcal P}_{(i,j)}(T_5) = ({\mathcal F}_i \circ {\mathcal F}_j \circ {\mathcal F}_i \circ {\mathcal F}_j \circ {\mathcal F}_i)(T_5)$ on the $5$-gon.
}
\end{figure}

\begin{proof}
Figure \ref{fig_Transposition} illustrates that $({\mathcal F}_i \circ {\mathcal F}_j \circ {\mathcal F}_i \circ {\mathcal F}_j \circ {\mathcal F}_i)$ amounts to the transposition of the labels $i$ and $j$ when performed on a triangulation of the $5$-gon. This proves our result since the other triangulations of the $5$-gon are similar to this one up to rotation. In particular, this result is true when dealing with a convex $n$-gon.
\end{proof}

More generally, this result illustrates that the flip can be used to change the connectivity of a triangulation, but also to move its edges. This remark is the starting point of a process to generate sequences of flips that are equivalent when performed on a triangulation, and to develop an algorithm to reduce sequences.
%--------------

\subsection{Equivalence between sequences of edge flips}

We first introduce two equivalence relations between sequences of edge flips performed on a labelled triangulation $T$. 

Let $\phi_1$ and $\phi_2$ be two sequences of edge flips. $\phi_1(T)$ and $\phi_2(T)$ are two triangulations of the same set of vertices.
\begin{itemize}
\item $\phi_1$ is \textit{strongly equivalent to}  $\phi_2$ when performed on $T$, if $\phi_1(T)=\phi_2(T)$,
\item $\phi_1$ is \textit{weakly equivalent to} $\phi_2$ when performed on $T$, if $\phi_1(T)$ and $\phi_2(T)$ correspond to the same triangulation up to a permutation of the edge labels;  we denote $\phi_1(T) \simeq \phi_2(T)$ in that case. 
\end{itemize}

In the following, we are interested in generating sequences that are either strongly or weakly equivalent to a given sequence $\phi$, when performed on $T$.  Let us observe that the strong equivalence implies weak equivalence and that a minimal sequence of edge flips between unlabelled triangulations $T$ and $\phi(T)$ is weakly equivalent to $\phi$ on $T$.

%-------------------------

\section{Generation of equivalent sequences of edge flips}
\label{sec_Generation}

\subsection{Introduction of three moves \label{sec_moves}}

The three moves we introduce are directly derived from the properties of the flip operation performed on a triangulation with labelled edges. They can be used to generate sequences that are either strongly or weakly equivalent to an input sequence $\phi$, when performed on an input triangulation $T$ $\in$ ${\mathcal T}_{lab}$. Furthermore, if $T$ is a triangulation of a convex $n$-gon, we will show that one can generate any sequence that is weakly equivalent to $\phi$ on $T$ using only these three moves. In the following, all the sequences we consider are composed of flippable edges, and the operations we introduce transform a sequence of flippable edges into another one where the flips remain valid.\\

\textbf{$1^{st}$ move (exploiting commutativity):} 
If there exists two sequences $\mu$ and $\nu$ such that $\phi = (\mu \circ {\mathcal F}_i \circ {\mathcal F}_{j} \circ \nu)$ and that $|supp(i, \nu(T)) \bigcap supp(j, \nu(T))| = 0$, then the sequence $(\mu \circ {\mathcal F}_j \circ {\mathcal F}_{i} \circ \nu)$ is strongly equivalent to $\phi$ on $T$. Therefore, if the sequence $\phi$ applied to $T$ contains flips that can be commuted, this move can be used to generate new sequences that are strongly equivalent to $\phi$. In the following, if two sequences can be transformed one into another, using only this first move, we say that they are \textit{strongly equivalent by commutativity}.\\

\textbf{$2^{nd}$ move (exploiting involution):} \\ 
We have the property that $({\mathcal F}_i \circ {\mathcal F}_i) (T) =T$ for any flippable edge $i$ in $T$. Therefore, one can generate strongly equivalent sequences by inserting $({\mathcal F}_i \circ {\mathcal F}_i)$ wherever $i$ is flippable in the intermediate triangulations produced by $\phi$ (expanding version of the $2^{nd}$ move). Conversely, if $\phi$ contains several occurrences of ${\mathcal F}_i$, and that there exists two sequences $\mu$ and $\nu$ such that $\phi$ is strongly equivalent by commutativity to the sequence $(\mu \circ {\mathcal F}_i \circ {\mathcal F}_i \circ \nu)$ when performed on $T$, then the first two moves can be used to generate the sequence $(\mu \circ \nu)$ which is strongly equivalent to $\phi$ but contains two fewer edge flips (reducing version of the $2^{nd}$ move).\\ 

\textbf{$3^{rd}$ move (exploiting transposition of labels):}\\

\hspace*{-5.4mm}\textbf{Proposition 2.}
\label{prop:2}
If there exists $\mu$ and $\nu$ such that $\phi=(\mu \circ {\mathcal F}_j \circ {\mathcal F}_i \circ {\mathcal F}_j \circ \nu)$ with $|supp(i, \nu(T)) \bigcap supp(j,\nu(T)) |=1$ and that $i$ and $j$ are also flippable in $\nu(T)$ and $({\mathcal F}_i \circ \nu)(T)$ respectively, then $(\mu \circ {\mathcal F}_j \circ {\mathcal F}_i \circ {\mathcal F}_j \circ \nu) (T) = (\mu \circ P_{(i,j)} \circ {\mathcal F}_j \circ {\mathcal F}_i \circ \nu) (T)$. 

\begin{figure}[h]
\centering
\includegraphics[scale=0.4]{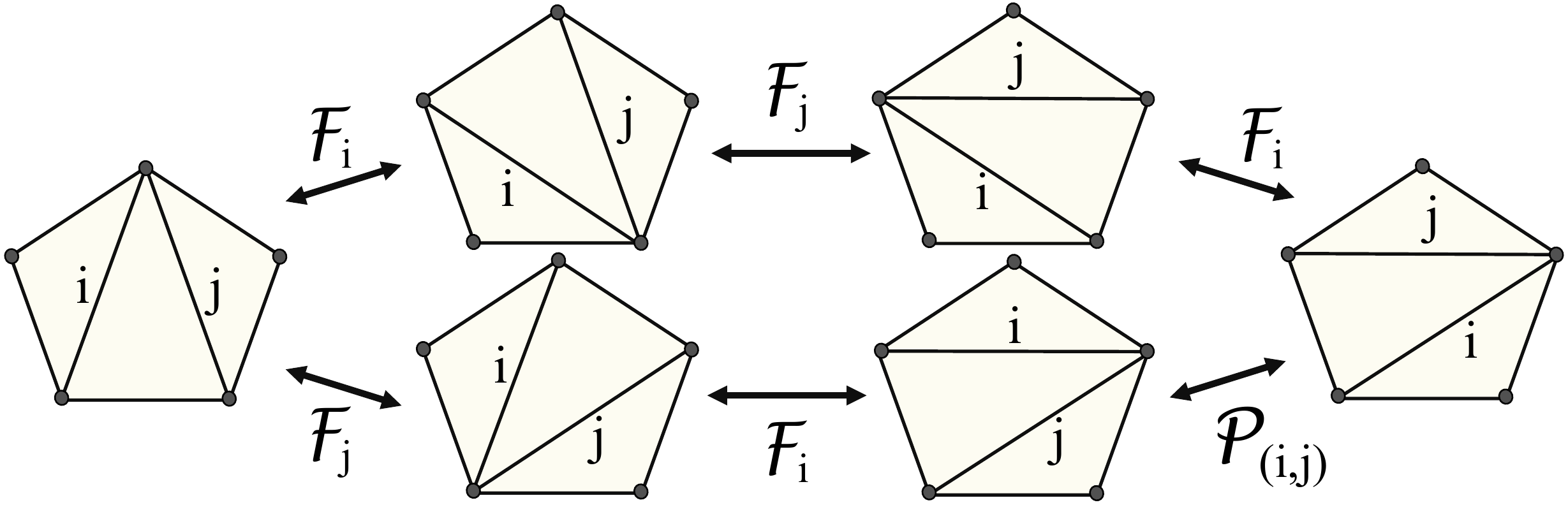}
\caption{\label{fig_pentagon}
In the triangulated pentagon, $({\mathcal F}_i \circ {\mathcal F}_j \circ {\mathcal F}_i) (T) = ({\mathcal P}_{(i, j)} \circ {\mathcal F}_i \circ {\mathcal F}_j) (T)$
}
\end{figure}

\begin{proof}
Since $|supp(i, \nu(T)) \bigcap supp(j, \nu(T))|=1$ and we can flip $i$ in $\nu(T)$ and $j$ in $({\mathcal F}_i \circ \nu)(T)$, the local configuration of the two supports is a $5$-gon where it is possible to flip these two inner edges successively. Let us consider $T'$ $=({\mathcal F}_j \circ {\mathcal F}_i \circ \nu)(T)$. Following the same reasoning as for Proposition 1, we have $({\mathcal F}_j \circ {\mathcal F}_i \circ {\mathcal F}_j \circ {\mathcal F}_i \circ {\mathcal F}_j) (T') = P_{(i,j)}(T')$.
We now replace $T'$ with $({\mathcal F}_j \circ {\mathcal F}_i \circ \nu)(T)$ in this equality, which amounts to $({\mathcal F}_j \circ {\mathcal F}_i \circ {\mathcal F}_j \circ {\mathcal F}_i \circ {\mathcal F}_j \circ {\mathcal F}_j \circ {\mathcal F}_i \circ \nu)(T) =  (P_{(i,j)} \circ {\mathcal F}_j \circ {\mathcal F}_i \circ \nu)(T)$ i.e. $({\mathcal F}_j \circ {\mathcal F}_i \circ {\mathcal F}_j \circ \nu) (T)=(P_{(i,j)} \circ {\mathcal F}_j \circ {\mathcal F}_i \circ \nu) (T)$ (see Figure \ref{fig_pentagon}).
\end{proof}

Let $P_{(i,j)}(\mu)$ denote the sequence $\mu$ in which every occurrence of ${\mathcal F}_i$ has been replaced with ${\mathcal F}_j$ and conversely.\\

The above property implies that $\phi=(\mu \circ {\mathcal F}_j \circ {\mathcal F}_i \circ {\mathcal F}_j \circ \nu)$ is weakly equivalent to $(P_{(i,j)} (\mu) \circ {\mathcal F}_j \circ {\mathcal F}_i \circ \nu)$ when performed on $T$ (reducing version of the $3^{rd}$ move corresponding to one flip fewer). \\

Conversely, if there exists $\mu$ and $\nu$ such that $\phi=(\mu \circ {\mathcal F}_j \circ {\mathcal F}_i \circ \nu)$ with $|supp(i, \nu(T)) \bigcap supp(j,\nu(T)) |=1$, that $j$ is flippable in $\nu(T)$ and $({\mathcal F}_i \circ {\mathcal F}_j \circ \nu)(T)$, and $i$ is flippable in $( {\mathcal F}_j \circ \nu)(T)$, then $\phi=(\mu \circ {\mathcal F}_j \circ {\mathcal F}_i \circ \nu)$ is weakly equivalent to $(P_{(i,j)} (\mu) \circ {\mathcal F}_j \circ {\mathcal F}_i \circ {\mathcal F}_j \circ \nu)$ when performed on $T$ (expanding version of the $3^{rd}$ move corresponding to one more flip).

%--------------

\subsection{Particular case of the convex $n$-gon}

When dealing with triangulations of a convex $n$-gon, any inner edge is flippable, whatever its support. Furthermore, we have the following result, regarding the equivalent sequences that can be generated with the three moves introduced in the previous section:

\hspace*{-5.4mm}\textbf{Theorem 1.} 
\label{theorem1}
Let $T \in {\mathcal T}_{lab}$ be a labelled triangulation of a convex $n$-gon and $\phi_1$ a sequence of edge flips. The three moves exploiting the commutativity and the involution of edge flips, in combination with the transposition of labels, can be used to generate any sequence of edge flips that is weakly equivalent to  $\phi_1$ on $T$. This means that the three moves are a generating set of the equivalence class associated with $\phi_1$.

\begin{figure}[h] 
\centering
\includegraphics[scale=0.6]{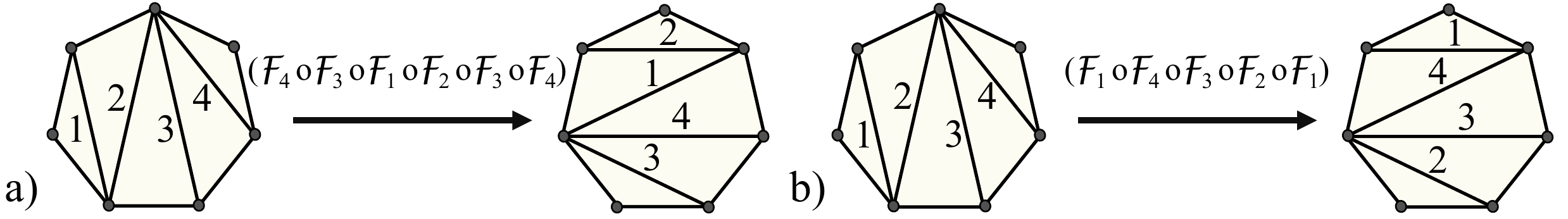}
\caption{\label{fig_ExempleSeq}
a) and b) correspond to two weakly equivalent sequences. 
}
\end{figure}

The proof of Theorem \ref{theorem1} can be found in the appendix. We believe that it constitutes a step forward in the understanding of edge flips, but the problem of exhibiting the minimal sequences remains complicated. Let us consider the example presented in Figure \ref{fig_ExempleSeq} that illustrates how the transformation of the initial sequence in $a)$ into the minimal sequence in $b)$ cannot easily be obtained by a direct algorithm :\\

\hspace*{-5.7mm}$({\mathcal F}_4 \circ {\mathcal F}_3 \circ \underline{{\mathcal F}_1 \circ {\mathcal F}_2} \circ {\mathcal F}_3 \circ {\mathcal F}_4)(T) = (P_{(1,2)} \circ {\mathcal F}_4 \circ {\mathcal F}_3 \circ {\mathcal F}_1 \circ {\mathcal F}_2 \circ {\mathcal F}_1 \circ {\mathcal F}_3 \circ {\mathcal F}_4)(T)$\\% (direct expansion)\\
\hspace*{48.5mm} $= (P_{(1,2)} \circ {\mathcal F}_4 \circ {\mathcal F}_1 \circ \underline{{\mathcal F}_3 \circ {\mathcal F}_2 \circ {\mathcal F}_3} \circ {\mathcal F}_1 \circ {\mathcal F}_4)(T)$\\
\hspace*{48.5mm} $= (P_{(1,2)} \circ P_{(2,3)} \circ {\mathcal F}_4 \circ {\mathcal F}_1 \circ {\mathcal F}_3 \circ {\mathcal F}_2 \circ {\mathcal F}_1 \circ {\mathcal F}_4)(T)$\\% (direct reduction)\\
\hspace*{48.5mm} $= (P_{(1,2)} \circ P_{(2,3)} \circ {\mathcal F}_1 \circ \underline{{\mathcal F}_4 \circ {\mathcal F}_3 \circ {\mathcal F}_4} \circ {\mathcal F}_2 \circ {\mathcal F}_1)(T)$\\
\hspace*{48.5mm} $= (P_{(1,2)} \circ P_{(2,3)} \circ P_{(3,4)} \circ {\mathcal F}_1 \circ {\mathcal F}_4 \circ {\mathcal F}_3 \circ {\mathcal F}_2 \circ {\mathcal F}_1)(T)$\\% (direct reduction)\\
\hspace*{48.5mm} $\simeq ({\mathcal F}_1 \circ {\mathcal F}_4 \circ {\mathcal F}_3 \circ {\mathcal F}_2 \circ {\mathcal F}_1)(T)$

%-------

\subsection{General case}

When working within a class of triangulations that is broader than the triangulations of the convex $n$-gon (possibly with internal vertices or even with concavities on boundaries, if in a geometric setting), the previous result probably does not directly generalize and we cannot ensure that a minimal sequence of flips could be reached by using only the three moves in their expanding and reducing versions.

%-----------------------------------------------

\section{Reducing sequences of edge flips}
\label{sec_reduction}

Based on the previous results and on the three moves introduced in section \ref{sec_moves}, we have devised an efficient algorithm for determining a reduced sequence that is weakly equivalent to an input sequence. It provides a heuristic alternative to the usually NP-complete problem of generating a sequence of minimal length between two triangulations. This algorithm illustrates the possibilities offered by our results. It can be used in a purely combinatorial setting or in combination with geometric intersection free predicates.
The reduced sequence obtained is locally minimal with regards to the other sequences that could be reached using any single occurrence of the expanding or reducing moves 2 and 3, in combination with the first move, but it is not necessarily a globally shortest edge flip sequence.

\subsection{Algorithm}

Given a triangulation $T$ and a sequence $\phi$ with $f$ flips, the algorithm attempts to eliminate each of the edge flips, considered one after the other, into a constructive order.

\subsubsection{Attempt to reduce $\phi$ by moving the flip ${\mathcal F}_i$ within $\phi$:}

Given an occurrence of ${\mathcal F}_i$ in $\phi$, we investigate whether the sequence $\phi$ contains another occurrence of ${\mathcal F}_i$ among the following flips. This requires a search among at most $n-1$ elements, with a complexity in $O(n)$.\\

If there is no occurrence of ${\mathcal F}_i$ among the following flips in $\phi$, we say that ${\mathcal F}_i$ cannot be removed and nothing is done. Otherwise, we denote ${\mathcal F}_i'$ the next occurrence of ${\mathcal F}_i$ in $\phi$, and we try to bring the two edge flips together in $\phi$: ${\mathcal F}_i$ is moved to the left, towards ${\mathcal F}_i'$, as long as ${\mathcal F}_i$ can be commuted with the next edge flip on the corresponding intermediate triangulation. Similarly, ${\mathcal F}_i'$ is moved to the right, towards ${\mathcal F}_i$, as long as ${\mathcal F}_i$ can be commuted with the previous edge flip on the corresponding intermediate triangulation. A move to the left denotes a move amongst the flips that were initially planned to be performed afterwards. Similarly, a move to the right is a move amongst the flips that were planned to be performed beforehand.

More formally, let $\phi_{ev}$ denote the evolving sequence of flips which is initialized to $\phi$. At each step of the displacement of ${\mathcal F}_i$, let $\mu_1$, $\mu_2$ and $\mu_3$ denote the three sequences such that $\phi_{ev}=(\mu_3 \circ {\mathcal F}_i' \circ \mu_2 \circ {\mathcal F}_j \circ {\mathcal F}_i \circ \mu_1)$. If we have $|supp(i, \mu_1(T)) \bigcap supp(j,\mu_1(T)) |=0$ and the edges can be flipped, ${\mathcal F}_i$ is moved to the left ; $\phi_{ev}$ is updated to $(\mu_3 \circ {\mathcal F}_i' \circ \mu_2 \circ {\mathcal F}_i \circ {\mathcal F}_j \circ \mu_1)$ and $\mu_1(T)$ is updated to $({\mathcal F}_j \circ \mu_1)(T)$. The displacement is stopped otherwise. The check of the independence between the supports of $i$ and $j$ is performed on $\mu_1(T)$, which means that we update $\mu_1(T)$ throughout the displacement of ${\mathcal F}_i$. It should be noted that each single displacement on the left corresponds to a flip in $\mu_1(T)$, and thus is performed in constant time. The total cost of displacing ${\mathcal F}_i$ on the left is $\Theta(l)$ with $l \leq f$.

Afterwards, we similarly displace ${\mathcal F}_i'$ on the right towards ${\mathcal F}_i$ as long as it is possible and the two occurrences of ${\mathcal F}_i$ do not cross. Each single displacement on the right is checked and performed in constant time, except for the first displacement of ${\mathcal F}_i'$, which accounts for $\Theta(k)$, where $k$ is the number of flips between the two occurrences of ${\mathcal F}_i$ ($k \leq f-l$). Therefore, the total cost of displacing ${\mathcal F}_i'$ on the right is $\Theta(k)$.

When the displacement of ${\mathcal F}_i'$ and ${\mathcal F}_i$ is over, three cases can be encountered:
\begin{itemize}
\item If $\phi_{ev}=(\nu \circ {\mathcal F}_i' \circ {\mathcal F}_i \circ \mu)$, then $\phi_{ev}$ is reduced to $(\nu \circ \mu)$ (two removal of ${\mathcal F}_i$, that is two edge flips fewer in constant time)
\item If $\phi_{ev}=(\nu \circ {\mathcal F}_i' \circ {\mathcal F}_j \circ {\mathcal F}_i \circ \mu)$ and $|supp(i, \mu(T)) \bigcap supp(j,\mu(T)) |=1$ with each edge $i$ and $j$ flippable within the sequence $({\mathcal F}_i \circ {\mathcal F}_j)$ performed on $\mu(T)$, then $\phi_{ev}$ is reduced to $(T_{(i,j)}(\nu) \circ {\mathcal F}_i \circ {\mathcal F}_j \circ \mu)$ (one removal of ${\mathcal F}_i$, that is one flip fewer in $O(n)$, since the sequence $\nu$ which may be transformed contains fewer than $f$ edge flips)
\item If neither of the previous cases is encountered, we say that $\phi$ cannot directly be reduced using ${\mathcal F}_i$. The first occurrence of ${\mathcal F}_i$ is reset to its initial position which is performed in constant time. %we let the two occurences of ${\mathcal F}_i$ in their displaced positions.
\end{itemize}

In the end, the cost of an attempt to reduce $\phi$ by moving the flip ${\mathcal F}_i$ within $\phi$ is $O(f)$ where $f$ is the number of edge flips in $\phi$.

%-------------------

\subsubsection{Algorithm for reducing sequences of edge flip:} 

The algorithm iterates over all the edge flips in $\phi$, from the beginning to the end of the sequence (right to left), until it finds a flip ${\mathcal F}_i$ whose displacement reduces $\phi$. The removal of an occurrence of ${\mathcal F}_i$ makes it necessary to reconsider a previously checked edge flip, in order to reduce the updated sequence. Therefore, after each reduction of $\phi$, the iteration is restarted from the beginning of the updated $\phi$.\\

After each reduction, the algorithm iterates over a subset of the $l \leq f$ remaining edges in $\phi_{ev}$ until it identifies the next reduction, if any. The complexity of finding the next reduction is $O(l^2)$. If no more edge can lead to a sequence reduction, the algorithm iterates over the $l \leq f$ edges of $\phi_{ev}$ to search for a last attempt at reduction with each remaining edge ($O(l^2)$ complexity). The sequence is stricly decreased in length after each reduction, so that there can be no more than $f$ reductions. Therefore the overall complexity of the algorithm is $O(f^3)$. There are several variations around this basic version of the algorithm, but they do not affect the overall complexity of the algorithm.

It can be noted that when applied to a sequence that is nearly optimal, the algorithm performs in $O(f^2)$ or even better ($O(f)$) if most of the edge flips are present only once and if we trigger the displacement of a couple of flips only when identifying a second occurrence of a flip. The algorithm reduces the number of edge flips in a sequence and exhibits a local minimum in terms of the length of the sequence. If it is possible to reduce the sequence further, we would have to go through temporary sequences whose length is longer. Ideas in this direction can be suggested by Theorem 1, but their integration in an effective algorithm is not that simple.

\begin{table}[h]
\scriptsize
\hspace*{2,4mm}
\begin{tabular}{|c||c|c|c||c|c|c||c|c|c||c|}
\hline
&\multicolumn{3}{c||}{ }	&\multicolumn{3}{c||}{ } &\multicolumn{4}{c|}{ }\\
&\multicolumn{3}{c||}{\textbf{Convex $n$-gon}}	&\multicolumn{3}{c||}{\textbf{Geometrical Setting}} &\multicolumn{4}{c|}{\textbf{Combinatorial Setting}}\\
&\multicolumn{3}{c||}{ }	&\multicolumn{3}{c||}{ } &\multicolumn{4}{c|}{ }\\
\hline
\textbf{Nb of edges}	&2 000	&10 000	&30 000	&2 000	&10 000	& 30 000	&2 000	&10 000 	&30 000	&100 000\\
\hline
\hline
\textbf{Initial}	&	&	&	&		&		&	&		&	& 	&\\
\textbf{size}	&200	&500	&2 000	&200		&500		&2 000	&200		&500	&2 000 	&10 000\\
($r$=1.1)	&	&	&	&		&		&	&		&	& 	&\\
\hline
\textbf{Gain}				&27		&64	&216		&2			&7			&70		&8		&31		&157	&1 043\\
(\%)				&(13) 		&(12)		&(10)		&(1)			&(1)			&(3)		&(4)		&(6)		&(7 )	&(10)\\
\hline
%&	&	&	&	&	&	&	&	& \\
\textbf{Time}				&1ms 	&13ms 	&76ms 		&1ms 	&4ms 	&31ms 	&5ms 	&24ms 	&541ms 	&2s\\
%&	&	&	&	&	&	&	&	& \\
\hline
\hline
\textbf{Initial}	&	&	&	&		&		&	&		&	& 	&\\
\textbf{size}&2 000	&6 000	&10 000		&2 000		&6 000	&10 000		&2 000	&6 000	&10 000	&20 000\\
($r$=2)	&	&	&	&		&		&	&		&	& 	&\\
\hline
\textbf{Gain} 			&973	&2 950	&4 964		&596		&1 062		&2698		&854&2 288	&4 608	&9 152\\
(\%)			&(48)	&(49)	&(49)		&(29)		&(17)		&(26)		&(42)	&(38)	&(46)	&(45)\\
\hline
%&	&	&	&	&	&	&	&	& \\
\textbf{Time}				&276ms 		&7s 	&1min22s 	&101ms 	&588ms 	&1s 		&3s 		&57s 				&1m28s 	&5min36s\\
%&	&	&	&	&	&	&	&	& \\
\hline
\hline
\textbf{Initial}	&	&	&	&		&		&	&		&	& 	&\\
\textbf{size}	&6 000	&10 000		&20 000			&6 000	&10 000	&20 000	&6 000	&10 000			&20 000 	&30 000\\
($r$=10)	&	&	&	&		&		&	&		&	& 	&\\
\hline
\textbf{Gain}				&5 277	&8 666		&17 759		&4 212	&5 686	&10 353		&4 617		&5 804		&15 060	&22 530\\
(\%)				&(87)	&(86)		&(88)		&(70)	&(43)	&(51)		&(76)		&(58)		&(75)	&(75)\\
\hline
%&	&	&	&	&	&	&	&	& \\
\textbf{Time}				&14s 	&1min15s 	&3min53s 	&281ms 	&946ms 	&3s 			&8s 			&5min52s 	&20min14s 	&1h32min\\
%&	&	&	&	&	&	&	&	& \\
\hline
\end{tabular}
\caption{\label{tab_simplification}
Experimental results : in the geometrical and combinatorial settings, we have used the same flip sequences, ensuring that all edge flips verify the constraints of each setting. This helps emphasize the effect of the choice of a class on the reduced flip sequence. The last column corresponds to a sequence generated and reduced on a purely combinatorial case.
}
\end{table}

\subsection{Experimental results}

The experimental results in Table \ref{tab_simplification} were calculated on a PC with an Intel Core 2 Duo 2 GHz and 4 Gb of RAM. We worked within three different classes of triangulations: convex $n$-gon, straight-edge triangulations (geometric setting) of a sphere (but an underlying surface with a different genus could also be used) and a purely combinatorial setting where all the connectivity requirements prescribed in this paper are satisfied and each vertex is associated with a single topological disk (as in \cite{Espinas12}). The sequences were randomly generated such that the triangulation remainded valid after each flip and there is at least the flip of one neighbouring edge between two occurrences of the same edge flip. Flip sequences are characterized by their length $f$ as well as their redundancy $r$ which is the ratio of the sequence length to the number of different edges involved. In particular, if the sequence has at most one occurrence of each flip, $r$ is equal to 1. Also, $f(1 - 1/r)$ gives an estimate of the number of flips that could be removed by the simplification.

It is worthwhile noting that the observed complexity of the algorithm is directly related to the redundancy $r$~: linear in the length of the original sequence for $r$ around 1, it becomes quadratic or even cubic when $r$ grows. In practice, we observe on the experimental results presented in \ref{tab_simplification} that the overall complexity approximation performs in $O(f^{3-2/r})$. Likewise, the gain essentially depends on the triangulation class and the redundancy of the initial sequence rather than its length. The resulting lengths are comparable to the theoretical bounds on flip distances. In particular, it is instructive to see that the same flip sequence generated in a geometric setting can be reduced further when being handled within a combinatorial setting, where we benefit from more possible intermediate triangulations. This reduces the number of flips removed but also the algorithm's running time. 

To our knowledge, there are no other algorithms to reduce sequences of edge flips, and we cannot compare our results to others. However, we note that the reduced sequences applied on a convex $n$-gon, contains less than $2n-10$ flips, i.e. the length is less than the upper bound introduced by Pournin \cite{Pournin12}.

\begin{figure}[h]
\centering
\includegraphics[scale=0.55]{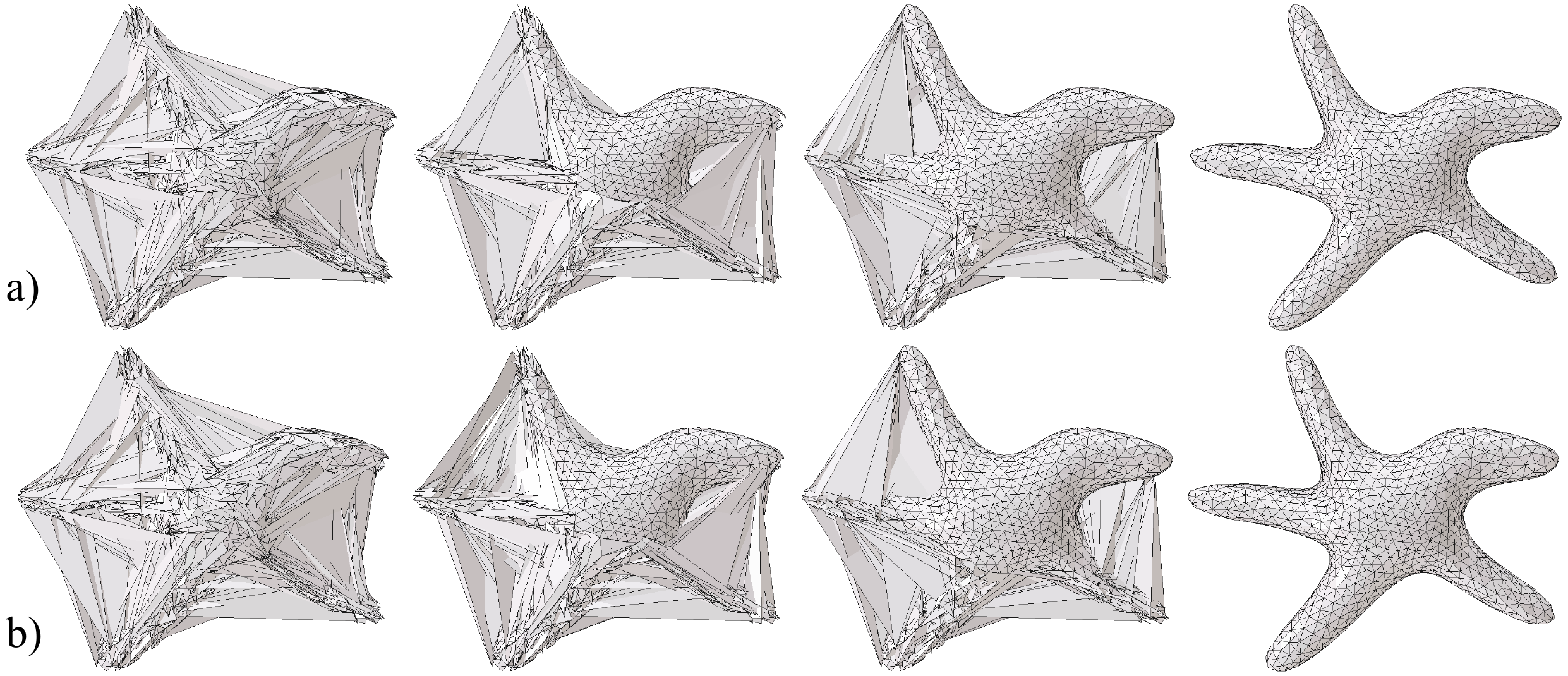}
\caption{\label{Reduction_exemple}
a) Progress of a sequence composed of 40 000 edge flips with redundancy $r=1.5$ (combinatorial setting). b) The reduced sequence composed of 29 942 edge flips. For each sequence, the figure illustrates the initial triangulation, the triangulation where $1/3$ of the sequence has been applied, then $2/3$ of the sequence, and the final triangulation.}

\end{figure}

Finally, we note that the progress of a triangulation on which we perform a sequence of flips is similar before and after the reduction (see Figure \ref{Reduction_exemple}).

%----------------------------------------------------

\section{Conclusion}

We have presented an efficient algorithm for reducing flip sequences based on the edge move study. We note that the gain is very close to the redundancy of the sequence, especially in the case of the $n$-gon, which demonstrates the practical value of the method for the efficient encoding of a triangulation connectivity. Indeed, several compression methods (\cite{devillers00,Valette09} for instance) are able to automatically recover much of the connectivity of a triangulation from its geometry information and could benefit, for the remaining connectivity, from a differential coding expressed as a simplified flip sequence. Another application could consist in using the labels to encode additional information. For example, we could assign colors to edges and transport them by successive permutations. We could also encode the transformation of a non-triangular graph by embedding it in a triangulation and marking the edges to be removed after the transformation.

%------------------------------------------------------------------------------

\newpage

\bibliographystyle{splncs}
\bibliography{egbibsample}

%------------------------------------------------------------------------------

\newpage

\section{Appendix : proof of the theorem}

\textbf{Theorem 1.} Let $T \in {\mathcal T}_{lab}$, a labelled triangulation of a convex $n$-gon and $\phi_1$ a sequence of edge flips. The three moves exploiting the commutativity and the involution of edge flips, in combination with the transpositions of labels, can be used to generate any sequences of edge flips that is weakly equivalent to  $\phi_1$ on $T$. This means that the three moves are a generating set of the equivalence class associated to $\phi_1$.

There may be other ways of proving this theorem, but we have chosen to demonstrate it with labelled edge flips, while introducing new results on triangulations with labelled edges.\\

\subsubsection{Preliminary definitions}

Let $\phi_1$ and $\phi_2$ denote two sequences of edge flips. For the needs of the demonstration of Theorem 1, we say that $\phi_2$ is a \textit{direct reduction} of $\phi_1$ if edge $i$ exists (and possibly an edge $j$) and two sequences $\mu$ and $\nu$ that satisfy one of the two following conditions:
\begin{itemize}
\renewcommand{\labelitemi}{$\bullet$}
\item $\phi_1$ is strongly equivalent by commutativity to $(\nu \circ {\mathcal F}_j \circ {\mathcal F}_i \circ {\mathcal F}_j \circ \mu)$ with $|supp(i, T) \bigcap supp(j,T) |=1$ and $\phi_2 = (T_{(i , j)}(\nu) \circ {\mathcal F}_j \circ {\mathcal F}_i \circ \mu)(T)$ (weak equivalence of $\phi_1$ and $\phi_2$ on $T$).
\item $\phi_1$ is strongly equivalent by commutativity to $(\nu \circ {\mathcal F}_i \circ {\mathcal F}_i \circ \mu)$ and $\phi_2 = (\nu \circ \mu)$ (strong equivalence of $\phi_1$ and $\phi_2$ on $T$).
\end{itemize}

We say that a sequence is \textit{reduced} if there is no direct reduction of it. By iterating direct reductions on an input sequence $\phi$, we can construct a \textit{monotonically reduced} version of it. There is weak equivalence between $\phi$ and its \textit{monotonic reductions}. It is noted that monotonically reduced versions of an input sequence are not unique and that monotonically reduced sequences are not necessarily minimal (see Figure \ref{fig_ExempleSeq} where a) corresponds to a non-minimal reduced sequence, since the sequence in b) contains one flip fewer).

In order to demonstrate theorem 1, we need the following lemmas: \\

\textbf{Lemma 1.}\label{lemma1}
Let $T \in {\mathcal T}_{lab}$ be a triangulation of a convex $n$-gon. If a sequence of flips $\phi$ contains no double occurrence of the same edge flip, then the sequence is minimal. Any other minimal sequence that is weakly equivalent to $\phi$ on $T$ is composed of the same edge flips (up to their order in the sequence).

\begin{proof}
The result is clear when $\phi$ is empty. 
We now consider the case where the length of $\phi$ is not zero, and there is no double occurrence of the same edge flip in $\phi$. If $\phi$ is not minimal, there exists a sequence $\phi_{min}$ that is weakly equivalent to $\phi$ on $T$ but whose length is strictly lower. Let ${\mathcal F}_i$ be a flip that is present in $\phi$ but not in $\phi_{min}$ (${\mathcal F}_i$ necessarily exists due to the length difference). Let $v_1$ and $v_2$ be the two vertices incident to $i$ in $T$ (but also in $\phi_{min}(T)$). Let $v_3$ and $v_4$ be the two vertices incident to $i$ in $\phi(T)$ (they are the two opposite vertices incident to the support of $i$ just before its flip). In the initial triangulation $T$, the edge $i$ splits the $n$-gon into two disconnected parts. They contain the vertices $v_3$ and $v_4$, respectively. This means that $v_3$ and $v_4$ are separated by $i$ and they cannot be connected without flipping $i$. However, $v_3$ and $v_4$ should be connected in $\phi_{min}(T)$, which contradicts the very existence of $\phi_{min}$ since $i$ is not flipped in $\phi_{min}$. Therefore $\phi$ is minimal.

Let us now assume that $\phi'$ is another minimal sequence such that $\phi'(T) \simeq \phi(T)$, but $\phi'$ does not flip the same edges as $\phi$. Let  ${\mathcal F}_i$ be a flip present in $\phi$ but not in $\phi'$. With the same reasoning as above, we deduce that $\phi$ and $\phi'$ cannot be equivalent on $T$, which is a contradiction.
\end{proof}

\begin{figure}[h]
\centering
\includegraphics[scale=0.55]{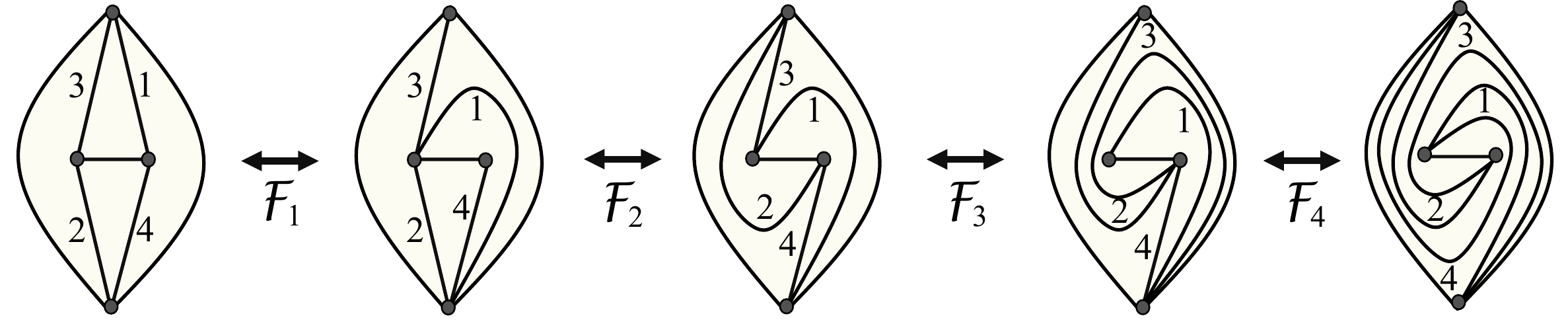}
\caption{\label{fig_Avec_Sommets_Interieurs}
The left and right triangulations are composed of the same edges and oriented faces but the labeling of the edges is different. Therefore, the sequence $({\mathcal F}_4 \circ {\mathcal F}_3 \circ {\mathcal F}_2 \circ {\mathcal F}_1)$ is weakly equivalent to the identity when performed on the left triangulation. This sequence is not minimal, but it contains no double occurrence of a same edge flip.
}
\end{figure}

We note that the Lemma \ref{lemma1} is not valid when there are internal vertices. Figure \ref{fig_Avec_Sommets_Interieurs} illustrates this point, with a counter-example of a non-minimal sequence without any redundancy in the edges that are flipped.

\textbf{Lemma 2.} Let $T \in {\mathcal T}_{lab}$ be a triangulation of a convex $n$-gon. If a minimal sequence of flips $\phi_1$ contains no double occurrence of the same edge flip, then any minimal sequence $\phi_2$ weakly equivalent to $\phi_1$ on $T$ is strongly equivalent to $\phi_1$ by commutativity.

\begin{figure}[h]
\centering
\includegraphics[scale=0.55]{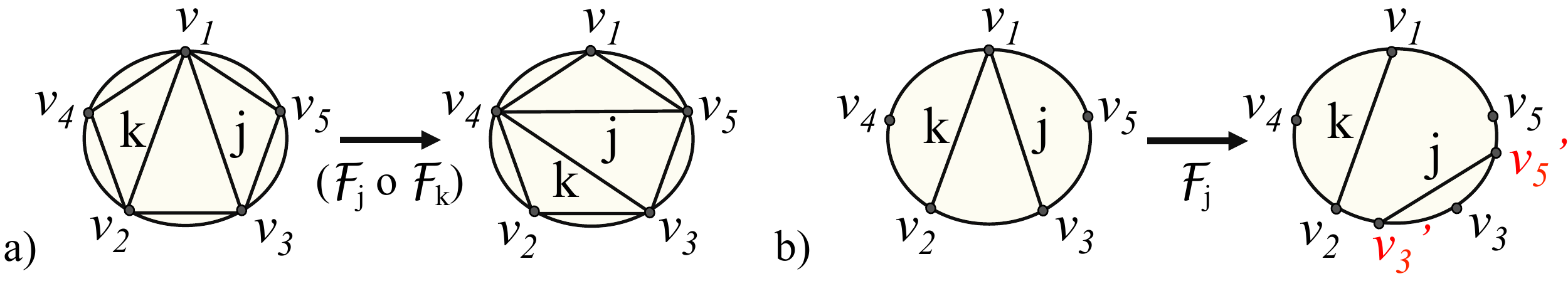}
\caption{\label{fig_DemoLemma}
a) Tracking the position of $k$ and $j$ in $({\mathcal F}_j \circ {\mathcal F}_k \circ \mu_2 \circ \nu)(T)$; b) evolution of $j$ in $({\mathcal F}_j \circ \nu)(T)$. It can be noted that $v_3'$ can coincide with $v_2$ and that the position of $v_5'$ with regards to $v_5$ is not specified. 
}
\end{figure}

\begin{proof}
Lemma 1 ensures that $\phi_1$ and $\phi_2$ are composed of the same edge flips. Let us seek to reorder, by commutativity, the flips of the sequence $\phi_2$ according to their position in $\phi_1$. If the first edge to be flipped in $\phi_1$ is ${\mathcal F}_i$, then we move ${\mathcal F}_i$ within $\phi_2$ towards the first position, as long as ${\mathcal F}_i$ can commute  with the edge flip that preceeds. If ${\mathcal F}_i$ can be moved to the first place, we repeat the process by moving the edge flip to be put in the position $p=2$, and so on. This operation is continued until one meets an edge flip ${\mathcal F}_j$ that cannot be moved to the desired position $p$ by commutativity. Therefore, there exist two sequences $\mu_1$ and $\nu$ with $|\nu|=p-1$, such that $\phi_1(T)=(\mu_1 \circ {\mathcal F}_j \circ \nu)(T)$, and there exist $\mu_2$ and $\mu_3$ such that $\phi_2(T)=(\mu_3 \circ {\mathcal F}_j \circ {\mathcal F}_k \circ \mu_2 \circ \nu)(T)$ and ${\mathcal F}_j$ and ${\mathcal F}_k$ do not commute on $(\mu_2 \circ \nu)(T)$. This means that $j$ and $k$ are incident to a common face (see Figure \ref{fig_DemoLemma} a). Let $v_1$ and $v_2$ be the two vertices incident to $k$ in the triangulation $(\mu_2 \circ \nu)(T)$. $k$ is incident to the faces $(v_1v_2v_3)$ and $(v_2v_1v_4)$. Without loss of generality, we can assume that the edge $j$ is incident to the vertices $v_1$ and $v_3$ and to the faces $(v_3v_1v_2)$ and $(v_1v_3v_5)$. 

Therefore, the edge $k$ is incident to $v_3$ and $v_4$ in $({\mathcal F}_j \circ {\mathcal F}_k \circ \mu_2 \circ \nu)(T)$, but also in $\phi_2(T)$ since it is flipped only once (see Figure \ref{fig_DemoLemma} a). It can be noted that $j$ is already incident to $v_1$ and $v_3$ in $\nu(T)$, since it is not flipped between $\nu(T)$ and $(\mu_2 \circ \nu)(T)$. After being flipped in $\phi_1$, $j$ splits the $n$-gon into two disconnected parts of $({\mathcal F}_j \circ \mu)(T)$. The first part contains $v_3$ and the second part contains all together $v_1$, $v_4$ and $k$ (see Figure \ref{fig_DemoLemma} b). This prevents the subsequent creation of an edge between $v_3$ and $v_4$ in $\phi_1(T)$, since $v_3$ and $v_4$ are separated by the already flipped $j$, which is a contradiction.
\end{proof}

\textbf{Lemma 3.} 
Let $T_F \in {\mathcal T}_{lab}$ be a triangulation of a convex $n$-gon that corresponds to a fan (all the edges are incident to a vertex $v_0$). We consider $\phi$ a sequence of edge flips on $T_F$. If $\phi$ is reduced on $T_F$ then $\phi$ is a minimal sequence. Furthermore, any other minimal sequence that is weakly equivalent to $\phi$ on $T_F$ is indeed strongly equivalent to $\phi$ by commutativity.\\

\begin{proof} 
Let us demonstrate that whatever the reduced sequence of flips in $T_F$, there is no double occurrence of the same edge flip. This can be done by induction on the number $n$ of vertices in the fan.\\

$\bullet$ The result is trivial when $n=3$ or $n=4$. Let us note that when $n=5$, reduced sequences are $Id$, ${\mathcal F}_1$, ${\mathcal F}_2$, ${\mathcal F}_1 \circ {\mathcal F}_2$ and ${\mathcal F}_2 \circ {\mathcal F}_1$ and none of them contains more than one occurrence of ${\mathcal F}_1$ and ${\mathcal F}_2$.\\

$\bullet$ Let us assume that the property is true on a convex $n$-gon, and let us extend it to a convex $(n+1)$-gon.

Let $\phi$ be a reduced sequence in a convex $(n+1)$-gon triangulated as a fan $T_F$ and $k$ ($0<k<n-2$) the first edge to be flipped in $\phi$. $v_0$ denotes the apex of the fan. There exists a sequence $\phi'$ such that $\phi=(\phi' \circ {\mathcal F}_k)$. $\phi'$ is reduced on ${\mathcal F}_k(T_F)$ since $\phi$ is reduced on $T_F$.

It can be observed that flipping an interior edge of a fan has the effect of decreasing the number of edges in this fan. In ${\mathcal F}_k(T_F)$, $k$ splits the $(n+1)$-gon into a single face and a convex $n$-gon triangulated as a fan of apex $v_0$ (see Figure \ref{fig_Flip_fan}). 

\begin{figure}[h]
\centering
\includegraphics[scale=0.5]{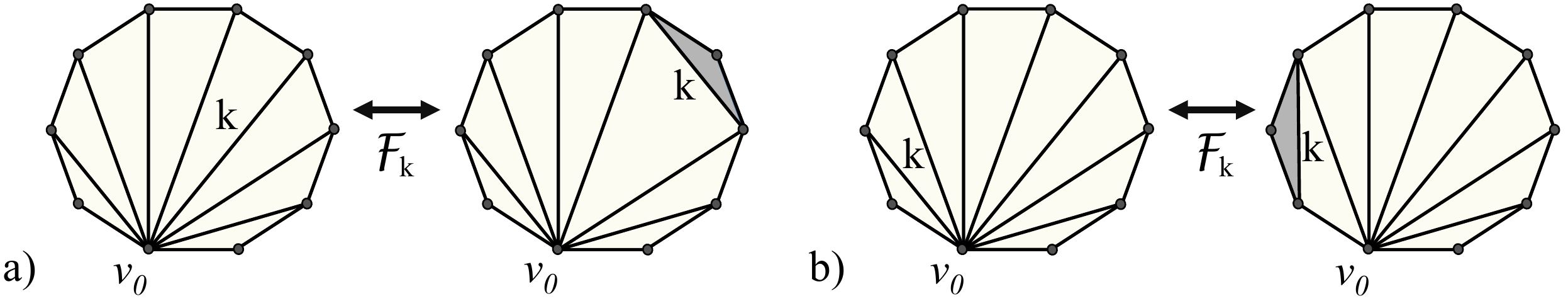}
\caption{\label{fig_Flip_fan}
For $n>0$, if $T_F$ is a convex $(n+1)$-gon with a fan triangulation, ${\mathcal F}_k(T_F)$ contains a convex $n$-gon with a fan triangulation.
}
\end{figure}

Let us assume that $\phi'$ contains an occurrence of ${\mathcal F}_k$.  Let $\mu$ and $\nu$ be two sequences such that $\phi=(\nu \circ {\mathcal F}_k \circ \mu \circ {\mathcal F}_k)$ and $\mu$ contains no occurrence of ${\mathcal F}_k$. ${\mathcal F}_k$ does not commute with $\mu$ since $\phi$ is reduced.

Therefore, there exists ${\mathcal F}_{k'}$ such that $\phi=(\nu \circ {\mathcal F}_k \circ \mu_2 \circ {\mathcal F}_{k'} \circ \mu_1 \circ {\mathcal F}_k)$ and ${\mathcal F}_{k'}$ is the first flip of the sequence that does not commute with the first occurrence of ${\mathcal F}_k$, ie  $({\mathcal F}_{k'} \circ \mu_1 \circ {\mathcal F}_k)(T_F) = ({\mathcal F}_{k'} \circ {\mathcal F}_k \circ \mu_1)(T_F)$ and $k$ and $k'$ are incident to a common face in $\mu_1(T_F)$.

The reduced sequence $\mu=(\mu_2 \circ {\mathcal F}_{k'} \circ \mu_1)$ is restricted to the fan included in ${\mathcal F}_k(T_F)$ since it contains no occurrence of  ${\mathcal F}_k$. By induction hypothesis, there is no double occurrence of the same flip in $(\mu_2 \circ {\mathcal F}_{k'} \circ \mu_1)$.

\begin{figure}[h]
\centering
\includegraphics[scale=0.55]{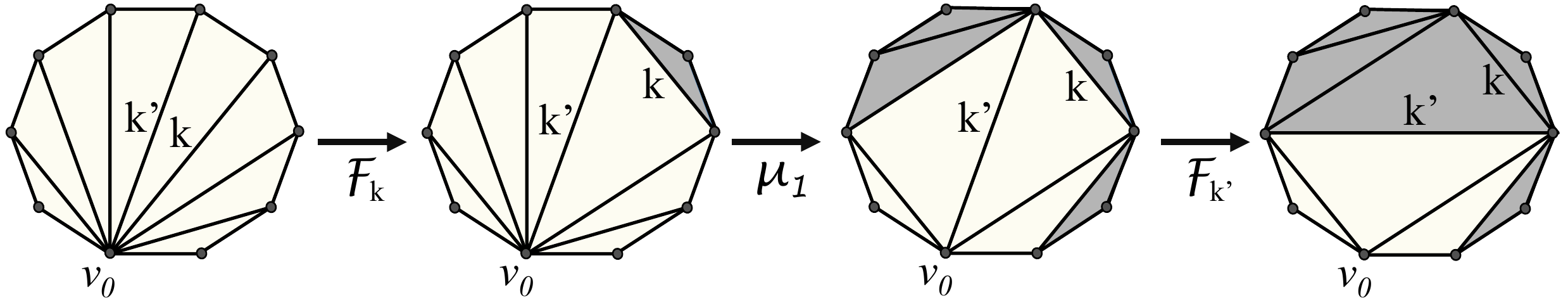}
\caption{\label{fig_Proposition}
$k'$ splits the triangulation $({\mathcal F}_{k'} \circ \mu_1 \circ {\mathcal F}_k)(T_F)$ into two disconnected parts: a $p$-gon triangulated as a fan and a part where $k$ is an inner edge.
}
\end{figure}

Furthermore, $k'$ splits $({\mathcal F}_{k'} \circ \mu_1 \circ {\mathcal F}_k)(T_F) = ({\mathcal F}_{k'} \circ {\mathcal F}_k \circ \mu_1)(T_F)$ into two disconnected parts (see Figure \ref{fig_Proposition}). Let $T_{F_p}$ denote the part that corresponds to a $p$-gon triangulated as a fan with $p < n$, and $T_k$ denotes the second part that contains $k$ as an inner edge. It can be noted that all the edges in $T_k$ that are different from $k$ have already been flipped in $\mu_1$. Therefore, the edges flipped in $\mu_2$ are located inside $T_{F_p}$ exclusively and ${\mathcal F}_k $ can be commuted with $\mu_2$ in $({\mathcal F}_{k'} \circ {\mathcal F}_k \circ \mu_1)(T_F)$. Thus $\phi(T_F)=(\nu  \circ \mu_2 \circ {\mathcal F}_k \circ {\mathcal F}_{k'} \circ {\mathcal F}_k \circ \mu_1)(T_F)$ with $|supp(k,\mu_1)(T_F)) \bigcap supp(k', \mu_1(T_F))| = 1$, which is not possible since $\phi$ is reduced. This ensures that $\phi'$ does not contain an occurrence of ${\mathcal F}_k$. Since $\phi'$ contains no double occurrence of the same edge split, by induction hypothesis, we have $\phi=(\phi' \circ {\mathcal F}_k)$ which satisfies the same property. Afterwards, Lemma 1 and 2 are used to close the demonstration.
\end{proof}

We can now prove Theorem $1$:
\begin{proof}
Let $\phi_1$ and $\phi_2$ two sequences of flips such that $\phi_1(T) \simeq \phi_2(T)$, we propose to show that one can generate a third sequence that is weakly equivalent to both $\phi_1$ and $\phi_2$ by using the three moves introduced in section \ref{sec_moves}.\\

Let $T_F$ be a fan triangulation of the input convex $n$-gon and $\mu$ a sequence of flips such that $\mu(T)=T_F$. We have $\phi_1(T)=(\phi_1 \circ \mu^{-1} \circ \mu)(T)$ and $\phi_2(T)=(\phi_2 \circ \mu^{-1} \circ \mu)(T)$. Let $\gamma_1$ be a reduced sequence of $(\phi_1 \circ \mu^{-1})$ and $\gamma_2$ be a reduced sequence of $(\phi_2 \circ \mu^{-1})$. Following Lemma 3, $\gamma_1$ and $\gamma_2$ are minimal and they are strongly equivalent by commutativity. Thus $\gamma_1$ can be generated from $\gamma_2$ using the move based on commutativity. This ensures that a combination of the three moves can be used on $\phi_1$ and $\phi_2$ to generate a sequence  $(\gamma_1 \circ \mu)$ that is weakly equivalent to both $\phi_1$ and $\phi_2$ on $T$. This also means that the three moves can be used to construct $\phi_2$ from $\phi_1$.
\end{proof}

\end{document}